\documentclass[aps,preprintnumbers,amsmath,amssymb,nofootinbib,superscriptaddress,notitlepage,10pt]{revtex4-1}

\usepackage{comment}
\usepackage{amssymb}
\usepackage{color}
\usepackage{rotating}
\usepackage{amsmath}
\usepackage{ulem}
\usepackage{overpic}
\usepackage{graphicx}
\usepackage{dcolumn}
\usepackage{graphicx}
\usepackage{bm}
\usepackage{chngpage}
\usepackage{multirow}
\usepackage{slashed}
\usepackage{indentfirst}
\usepackage{booktabs}
\usepackage{amssymb,amsfonts}
\usepackage{amsmath}
\usepackage{color}
\usepackage{hyperref}

\begin{document}

\title{Neutron drip line of $Z=9-11$ isotopic chains}
\author{Rong An}

\affiliation{School of Physics,
Beihang University, Beijing 100191, China}

\author{Guo-Fang Shen}
\affiliation{School of Physics,
Beihang University, Beijing 100191, China}

\author{Shi-Sheng Zhang}
\email[E-mail: ]{zss76@buaa.edu.cn}
\affiliation{School of Physics,
Beihang University, Beijing 100191, China}

\author{Li-Sheng Geng}
\email[E-mail: ]{lisheng.geng@buaa.edu.cn}
\affiliation{School of Physics,
Beihang University, Beijing 100191, China}
\affiliation{
Beijing Key Laboratory of Advanced Nuclear Materials and Physics, Beihang University, Beijing 100191, China}
\affiliation{Beijing Advanced Innovation Center for Big Data-based Precision Medicine, Beihang University, Beijing100191, China}
\affiliation{School of Physics and Microelectronics, Zhengzhou University, Zhengzhou, Henan 450001, China}


\begin{abstract}
  A recent experimental breakthrough identified the last bound neutron-rich nuclei in fluorine and neon isotopes. Based on this finding, we perform a theoretical study of $Z=9, 10, 11, 12$ isotopes in the relativistic mean field (RMF) model. The mean field parameters are assumed from the PK1 parameterization, and the pairing correlation is described by the particle number conservation BCS (FBCS) method recently formulated in the RMF model. We show that the FBCS approach plays an essential role in reproducing experimental results of fluorine and neon isotopes. Furthermore, we predict $^{39}$Na and $^{40}$Mg to be
  the last bound neutron-rich nuclei in sodium and magnesium isotopes.
\end{abstract}


\maketitle
\section{Introduction}
Properties of neutron-rich nuclei, in particular, the location of the neutron drip line,
play an important role not only in understanding nuclear stability with respect to the isospin
in hereto unexplored regions of the nuclear chart~\cite{Thoennessen_2013}, but also in
numerous related scientific issues of current interests. For instance, the r-process
nucleosynthesis of heavy elements in stellar evolution crucially depends on the
values of beta decay rates and neutron capture cross sections in neutron-rich
nuclei that do not exist in terrestrial conditions~\cite{Chen:1995nfm,KAJINO2019109,Li2019}.
The masses of neutron-rich nuclei impose stringent constraints on the equation
of state of neutron-rich nuclear matter, which is key to understanding the properties
of neutron stars and supernovae explosions~\cite{Geng:2005yu,Lattimer:2006xb,PhysRevC.88.024308,PhysRevC.88.061302,Hebeler:2013nza}.

New generation facilities, such as the Facility for Rare Isotope Beams (FRIB)
at Michigan State University and the Radioactive Isotope Beam Factory (RIBF) at RIKEN,
have helped in conducting studies of nuclei up to extreme isospin symmetry in the last two decades~\cite{BLANK2008403,RevModPhys.84.567,TANIHATA2013215,refId0}.
At the RIBF, the heaviest fluorine and neon isotopes were recently determined to be $^{31}$F and $^{34}$Ne~\cite{PhysRevLett.123.212501}, extending the neutron drip line
from $Z=8$~\cite{LANGEVIN198571,PhysRevC.41.937,PhysRevC.53.647,TARASOV199764,SAKURAI1999180}, determined twenty years ago, to $Z=10$.

\begin{table}[!hbp]
\begin{center}
\caption{Last bound neutron-rich nuclei with $Z=8-12$ predicted by various theoretical models in comparison with experimental data.}\label{tab1}
\begin{tabular}{lcc}
\hline\hline
 &  theory & experiment \\
\hline
~~~~~$Z=8$ & $^{24}$O~\cite{PhysRevC.60.054315,Wang:2014qqa},
$^{26}$O~\cite{PhysRevLett.116.102503}, $^{28}$O~\cite{Lalazissis:1997if,PhysRevC.88.024308,PhysRevC.88.061302,Goriely:2014qja,RevModPhys.41.S1,Xia:2017zka}
& $^{24}$O~\cite{LANGEVIN198571,TARASOV199764,PhysRevC.41.937,PhysRevC.53.647,Sakurai:1999tpf,Nakamura:2017xpt}\\
\hline
~~~~~$Z=9$ & $^{29}$F~\cite{Lalazissis:1997if,PhysRevC.88.024308,PhysRevC.88.061302,Goriely:2014qja,Wang:2014qqa}, $^{31}$F~\cite{PhysRevLett.122.052501,lalazissis2004},
$^{33}$F~\cite{Tanihata:1994wr} & $^{31}$F~\cite{Sakurai:1999tpf,PhysRevLett.123.212501}\\
\hline
~~~~~$Z=10$ & $^{30}$Ne~\cite{RevModPhys.41.S1},
$^{32}$Ne~\cite{PhysRevC.54.R2802}, $^{34}$Ne~\cite{Notani:2002ncs,Geng:2003wt,Wang:2014qqa,PhysRevLett.122.052501,PhysRevC.88.024308,PhysRevC.88.061302,
Goriely:2014qja,Tanihata:1994wr,PhysRevC.60.054315},
$^{38}$Ne~\cite{Zhou:2009sp},
$^{40}$Ne~\cite{Lalazissis:1997if},
$^{42}$Ne~\cite{Xia:2017zka}  & $^{34}$Ne~\cite{PhysRevLett.123.212501}\\
\hline
~~~~~$Z=11$ & $^{37}$Na~\cite{Geng:2003wt,Notani:2002ncs,Wang:2014qqa},
$^{39}$Na~\cite{RevModPhys.41.S1},
$^{45}$Na~\cite{Xia:2017zka,Meng:1998xq} & ${^{39}\mathrm{Na}}$(only one event)~\cite{PhysRevLett.123.212501} \\
\hline
~~~~~$Z=12$ & $^{40}$Mg~\cite{Baumann2007,PhysRevLett.122.052501,Zhi:2006wvp,Ren:1996jzt,TERASAKI1996371,TERASAKI1997706,Wang:2014qqa},
 $^{42}$Mg~\cite{PhysRevC.68.054312,PhysRevC.68.034311,PhysRevC.82.035804,Li:2012gv,RevModPhys.41.S1},
 $^{44}$Mg~\cite{MOLLER1995185},
 $^{46}$Mg~\cite{Li:2012gv,Zhou:2009sp,Xia:2017zka}&    \\
\hline\hline
\end{tabular}
\end{center}
\end{table}
Numerous theoretical studies on the location of the neutron drip line have been conducted.
Table~1 presents a partial list of various theoretical predictions in comparison
with the experimental data. Clearly, not all of the theoretical results agree with those
of the experiments and with each other. This is well understood, as the exact location
of the neutron drip line is sensitive to the details of the structure of nuclei, such as
the shell evolution, coupling between the continuum and bound states, deformation
effect, and three-body force. For example, the authors of Ref.~\cite{TANIHATA1995769}
argued that the neutron drip line is related to closing (sub)shell orbitals, and therefor
the drip line of fluorine and neon may be due to the closure of the $2p_{3/2}$
orbital, which results in formation of the drip line nuclei of $^{31}$F and $^{34}$Ne with $N=24$~\cite{PhysRevLett.123.212501}. In Refs.~\cite{PhysRevC.41.1147,MOTOBAYASHI19959,Motobayashi}, the extra stability of the
neutron-rich isotopes of fluorine and neon relative to oxygen isotopes were attributed to the emergence of
the island of inversion ($Z=10-12, N=20-22$, and their neighbors), where the ground states gain energy by strong
deformation. In Ref.~\cite{Otsuka:2009cs}, the three-body force was shown to be responsible for moving
the oxygen drip line from $^{28}$O to $^{24}$O.

At the mean field level, pairing correlations and continuum effects
play an important role in describing the ground-state properties of drip-line nuclei~\cite{Campi:1975mkn,Patra:1991xb,Starodubsky:1991ze,Starodubsky:1991he,Bertsch:1991zz,
PhysRevC.53.2809,Lalazissis:1997if,Bennaceur:1999tn,Grasso:2001hf,
Matsuo:2002gu,Hamamoto:2003ai,Tajima:2005hz,Yoshida:2006ji,Matsuo:2006iq}.
Pairing correlations are responsible for scattering nucleon pairs located at the single-particle level
below the Fermi surface into the level above with low orbital angular momentum.
Although the Bogoliubov method is arguably more appropriate to deal with the
pairing correlations in dripline nuclei~\cite{Dobaczewski:1983zc,PhysRevC.53.2809}, the BCS method, if treated properly, can offer a reasonable description~\cite{Sandulescu:1998dm,Sandulescu:2003jw,Geng:2003pk,Zhang:2014uca}. However, both methods have the drawback that the particle number is not conserved. Around the Fermi surface, only few
nucleons play an important role in determining the properties of neutron-rich nuclei, and the non-conservation of particle number may become particularly relevant. Therefore, the restoration of the particle number conservation is preferred, as shown in the present study.

Based on these considerations, in the present work,
we study fluorine, neon, sodium, and magnesium isotopes
in the relativistic mean field model and employ the
recently developed particle number conservation BCS (FBCS) approach to deal with pairing correlations~\cite{RongAn:114101}. We focus on the differences between the results obtained by the RMF+BCS and RMF+FBCS approaches to explore the relevance of the latter in the description of neutron-rich nuclei and in terms of predicting the location of the neutron drip line.

This paper is organized as follows. In Sec. 2,
we briefly describe the relativistic mean field model and the FBCS approach.
In Sec. 3, we present the results for fluorine, neon, sodium, and magnesium
 isotopes and discuss the differences between the BCS and FBCS approaches. A short
 summary and outlook are provided in Sec. 4.

\section{Theoretical model}
The relativistic mean field models or covariant density functional theories
have made remarkable progress in describing various nuclear physics
phenomena. Refs.~\cite{Walecka:1974qa,Reinhard:1989zi,Serot:1984ey,Ring:1996qi,Meng:2005jv,Vretenar:2005zz,Niksic:2011sg, Liang:2014dma, jie2016relativistic} provide further detail on this topic. In the present work, we take the more
conventional meson-exchange formulation, starting from the Lagrangian density expressed as Eq.~(\ref{eq1}), which contains the nucleon and the exchanged $\sigma$, $\rho$, $\omega$ meson, and the photon fields,
\begin{eqnarray}\label{eq1}
\mathcal{L}&=&\bar{\psi}[i\gamma^\mu\partial_\mu-M-g_\sigma\sigma
-\gamma^\mu(g_\omega\omega_\mu+g_\rho\vec
{\tau}\cdotp\vec{\rho}_{\mu}\\\nonumber
&+&e\frac{1-\tau_3}{2}A_\mu)
-\frac{f_\pi}{m_\pi}\gamma_5\gamma^\mu\partial_\mu
\overrightarrow{\pi}\cdotp\overrightarrow{\tau}]\psi\\\nonumber
&+&\frac{1}{2}\partial^\mu\sigma\partial_\mu\sigma-\frac{1}{2}m_\sigma^2\sigma^2
-\frac{1}{3}g_{2}\sigma^{3}-\frac{1}{4}g_{3}\sigma^{4}\\\nonumber
&-&\frac{1}{4}\Omega^{\mu\nu}\Omega_{\mu\nu}+\frac{1}{2}m_{\omega}^2\omega_\mu\omega^\mu
+\frac{1}{4}c_{3}(\omega^{\mu}\omega_{\mu})^{2}\\\nonumber
&-&\frac{1}{4}\vec{R}_{\mu\nu}\cdotp\vec{R}^{\mu\nu}+\frac{1}{2}m_\rho^2\vec{\rho}^\mu\cdotp\vec{\rho}_\mu
+\frac{1}{4}d_{3}(\vec{\rho}^{\mu}\vec{\rho}_{\mu})^{2}\\\nonumber
&-&\frac{1}{4}F^{\mu\nu}F_{\mu\nu},
\end{eqnarray}
where $M$ depicts the mass of the nucleon and $m_{\sigma}$, $m_{\omega}$, and $m_{\rho}$ are the masses of the $\sigma$, $\omega$, and $\rho$ mesons, respectively. Here $g_{\sigma}$, $g_{\omega}$, $g_{\rho}$, $g_2$, $g_3$, $c_3$, $d_3$, and $e^{2}/4\pi$ are coupling constants for the $\sigma$, $\omega$, $\rho$ mesons and photon. $\psi$ is the Dirac spinor for the nucleon.
The field tensors for vector mesons and the photon are
defined as $\Omega_{\mu\nu}=\partial_{\mu}\omega_{\nu}-\partial_{\nu}\omega_{\mu}$,
$\vec{R}_{\mu\nu}=\partial_{\mu}\vec{\rho}_{\nu}-\partial_{\nu}\vec{\rho}_{\mu}-g_{\rho}(\vec{\rho}_{\mu}\times\vec{\rho}_{\nu})$, and $F_{\mu\nu}=\partial_{\mu}A_{\nu}-\partial_{\nu}A_{\mu}$.

In the present study, we employ the parameter set PK1~\cite{PhysRevC.69.034319} for the mean-field effective interactions.
The PK1 parameter set is obtained by fitting to the masses of a number of spherical nuclei and the compression modulus, baryonic density at saturation, and asymmetry energy of nuclear matter. It could provide a  description of both the empirical properties of nuclear matter and the ground-state properties of finite nuclei better than most other parameter sets of the same group~\cite{PhysRevC.69.034319}. Because our main interest is to study the impact of the pairing correlation, particularly the particle number conservation, on the prediction of the dripline, we should only compare results obtained with the same mean-field effective force. Nevertheless, we performed studies using the NL3 parameter set, as in Ref.~\cite{RongAn:114101}, and our conclusion remains qualitatively unchanged. In the future, one may evaluate whether the impact of the particle number conservation on drip-line nuclei remains in the point-coupling version of the RMF model.

From the Lagrangian density, by employing the so-called no-sea and mean-field approximations,
the Dirac equation can be obtained for the nucleon, and the Klein-Gordon equations can be obtained for
the mesons. These equations can be solved self-consistently either in coordinate space or using the
basis expansion method. In the present study, we adopt the harmonic oscillator basis expansion method described in Refs.~\cite{Gambhir:1989mp,Ring:1997tc}, where the axial deformation degrees of freedom are taken into account.
In the numerical calculation, 12 shells are used to expand the fermion fields and 20 shells for the meson fields, which was found to be sufficient for the relatively small mass nuclei of this study~\cite{Geng:2003pk}.

To deal with pairing correlations, we employ the variational BCS approach~\cite{A.Bohr,P.Ring}. To restore the particle number
conservation, we adopt the variation after projection BCS method or the so-called FBCS method, which has been recently implemented in the RMF model~\cite{RongAn:114101}. The corresponding FBCS equation reads
\begin{eqnarray}
2(\widetilde{\varepsilon_{j}}+\Lambda_{j})\mu_{j}\nu_{j}+\Delta_{j}(\nu_{j}^{2}-\mu_{j}^{2})=0,
\end{eqnarray}
where the quantities $\tilde{\varepsilon}_{j}$, $\Lambda_{j}$, and $\Delta_{j}$ are defined as follows:
\begin{eqnarray}
\widetilde{\varepsilon_{j}}&=&(\varepsilon_{j}-G_{jj}\nu_{j}^{2})\frac{R_{1}^{1}}{R^{0}_{0}},\\\nonumber
\Delta_{j}&=&\sum_{k>0}G_{jk}\mu_{k}\nu_{k}\frac{R_{1}^{2}(j,k)}{R^{0}_{0}}(\nu_{j}^{2}-\mu_{j}^{2}),\\\nonumber
\Lambda_{j}&=&\sum_{k>0}(\varepsilon_{j}-\frac{1}{2}G_{kk}\nu_{k}^{2})\nu_{k}^{2}\frac{R^{0}_{0}(R_{2}^{2}
-R_{1}^{2})-R_{1}^{1}(R_{1}^{1}-R_{0}^{1})}{(R^{0}_{0})^{2}}\\\nonumber
&-&\frac{1}{2}\sum_{k_{1},k_{2}>0}G_{k_{1}k_{2}}\mu_{k_{1}}\nu_{k_{1}}\mu_{k_{2}}\nu_{k_{2}}\frac{R^{0}_{0}(R_{2}^{3}-R_{1}^{3})}{(R^{0}_{0})^{2}}\\\nonumber
&+&\frac{1}{2}\sum_{k_{1},k_{2}>0}G_{k_{1}k_{2}}\mu_{k_{1}}\nu_{k_{1}}\mu_{k_{2}}\nu_{k_{2}}\frac{R_{1}^{2}(R_{1}^{1}-R_{0}^{1})}{(R^{0}_{0})^{2}}.
\end{eqnarray}
In the above equations, $\epsilon_j$ is the single-particle energy,  $\nu_j^2$ ($\mu_j^2=1-\nu_j^2$) is the occupation probability of the $j$ orbit and its time reversal partner, and $G_{j_{1} j_{2}}=-\bar{V}_{j_{1}, \bar{j}_{1}, j_{2}, \bar{j}_{2}}$ is the pairing interaction matrix element with $V$ the pairing interaction, for which, in the present work, we adopt a density-independent delta interaction. The residuum integrals, $R_{0}^{0}$, $R_{0}^{1}$, $R_{1}^{1}$, $R_{1}^{2}$, $R_{2}^{2}$, $R_{1}^{3}$, and $R_{2}^{3}$, can be easily calculated, as explained in Ref.~\cite{RongAn:114101}.

\section{RESULTS AND DISCUSSIONS}
\begin{table}[htbp]
\begin{center}
\caption{Nuclei fitted to fix pairing interaction and corresponding pairing strength $V_{0}$ in units of $\mathrm{MeV}~\mathrm{fm}^{3}$.}\label{tab3}
\begin{tabular}{c|c|c|c|c}
\hline
\hline
 & $^{20-24}$F & $^{22-26}$Ne  & $^{26-30}$Na & $^{26-30}$Mg \\
 \hline
 RMF+BCS & $V_{0}=380$ & $V_{0}=580$ & $V_{0}=480$ & $V_{0}=480$ \\
 \hline
 RMF+FBCS & $V_{0}=380$ & $V_{0}=480$& $V_{0}=380$ & $V_{0}=380$ \\
\hline
\hline
\end{tabular}
\end{center}
\end{table}

To solve the RMF+BCS/FBCS equations, we apply a pairing window of $12$ MeV both
above and below the Fermi surface and adopt the density-independent contact delta interaction $V=-V_{0}\delta(\vec{r}_{1}-\vec{r}_{2})$ in the particle-particle channel~\cite{Geng:2003pk}.
The only free parameter in the particle-particle channel
is the pairing strength $V_{0}$, which is often adjusted to reproduce the odd-even mass staggerings. Here, we
use the three-point formula for such a purpose~\cite{P.Ring,A.Bohr}:
\begin{eqnarray}\label{oes}
\Delta&=&\frac{1}{2}[B(N-1,Z)-2B(N,Z)+B(N+1,Z)],\\\nonumber
\end{eqnarray}
where $B(N,Z)$ is the binding energy for a nucleus of neutron number $N$ and proton number $Z$. For each isotopic chain studied in either the BCS or the FBCS method, the pairing strength is fixed, as explained above, and it is provided in Table~2. We chose to fix the pairing strength by fitting to the odd-even staggerings of a few stable nuclei, as specified in Table~2, instead of fitting to the whole isotopic chain including the drip line nuclei. The reason is that our main purpose is to study the impact of the particle number conservation on the prediction of the drip line, but not to provide the best description of the existing data, which might require fine-tuning of the mean-field effective force and consideration of beyond mean field effects, such as configuration mixing and rotation corrections.

To determine the position of the drip line, we employ one- and two-neutron separation energies, given by Eqs.
(5) and (6), respectively, as follows:
\begin{equation}
S_{N}(N,Z)=B(N,Z)-B(N-1,Z),
\end{equation}
\begin{equation}
 S_{2N}(N,Z)=B(N,Z)-B(N-2,Z).
 \end{equation}

 The first unbound nucleus clearly has a negative $S_N$.
 For nuclei with an even neutron number, one should also verify $S_{2N}$ due to the pairing correlation, as $S_N$ might be negative. whereas $S_{2N}$ could be positive.

\begin{figure}[htbp]
   \includegraphics[scale=0.4]{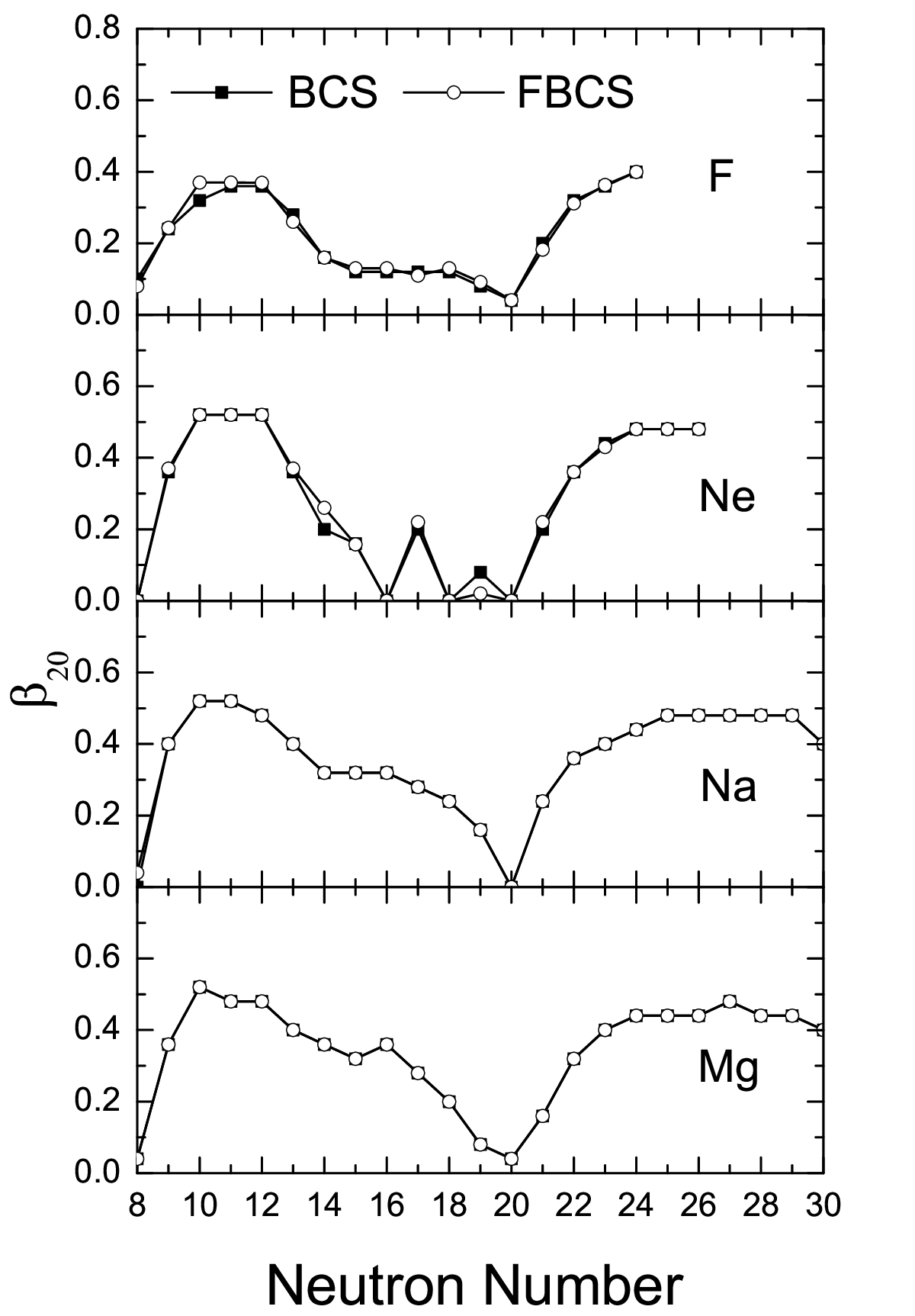}
     \includegraphics[scale=0.4]{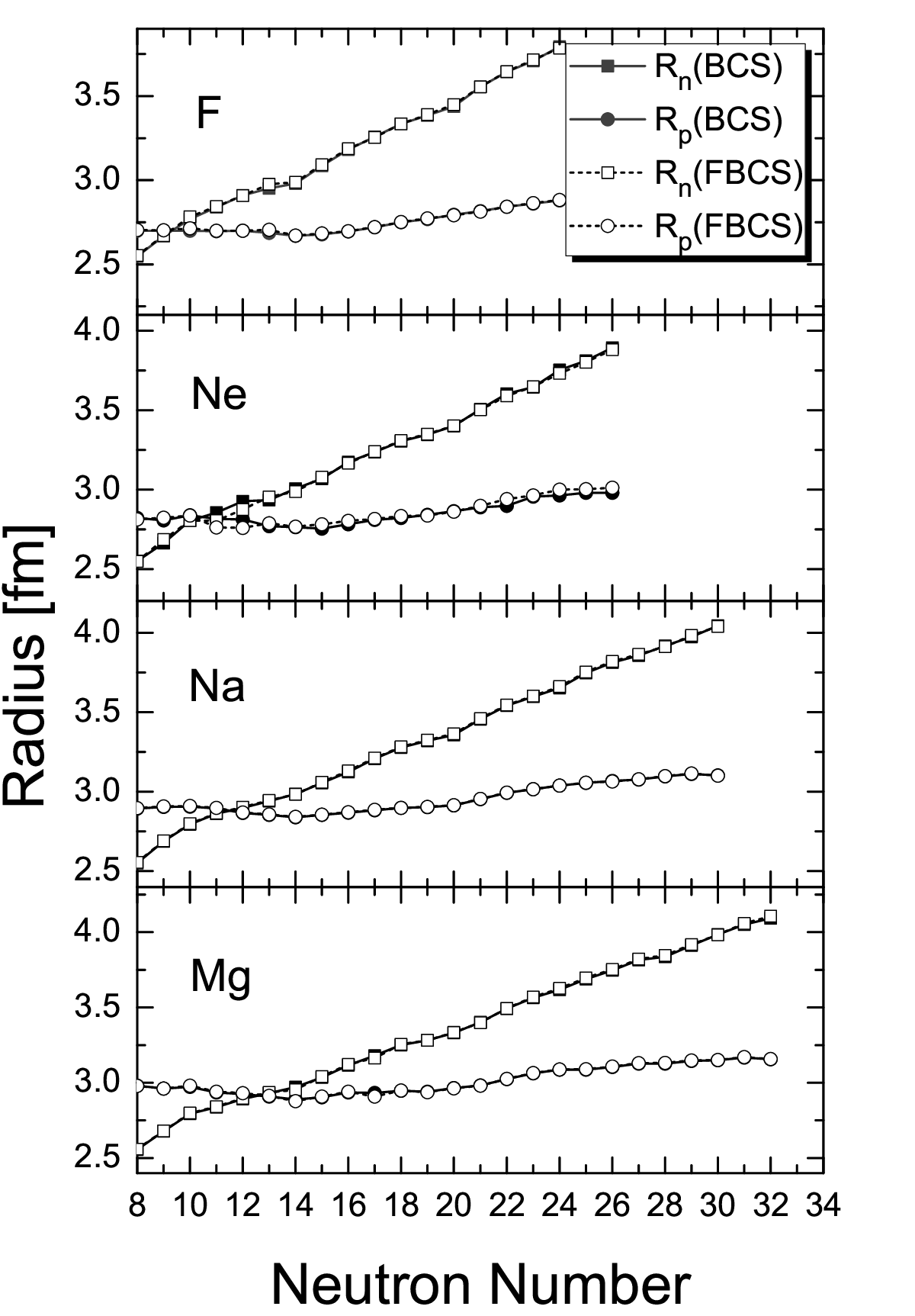}
    \caption{Quadrupole deformation parameters $\beta_{20}$ (left) and neutron ($R_n$) /proton ($R_p$) radii (right) of
    fluorine, neon, sodium, and magnesium isotopes obtained by RMF+FBCS and RMF+BCS approaches.\label{fig0}}
\end{figure}

In the following, we mainly concentrate on the odd-even staggerings, one-, and two-neutron separation energies,
as the quadrupole deformations and neutron/proton radii of the fluorine, neon, sodium, and magnesium isotopes are almost the same in the RMF+FBCS and RMF+BCS approaches, as shown in Fig.~\ref{fig0}. Clearly, most neutron-rich nuclei are deformed.~\footnote{The sudden increase in the $\beta_{20}$ of $^{27}$Ne is due to the fact that its binding energy is rather
soft with respect to its shape, as can be seen from an explicit constrained calculation. For such nuclei, to reliably determine its shape, configuration mixing effects need to be considered.} Therefore, one could imagine that spherical calculations may not be able to correctly predict/reproduce the drip line.

\begin{figure}[htbp]
 \begin{center}
  \setlength{\abovecaptionskip}{5pt}
  \begin{minipage}[c]{1.0\linewidth}
    \centering
   \includegraphics[scale=0.5]{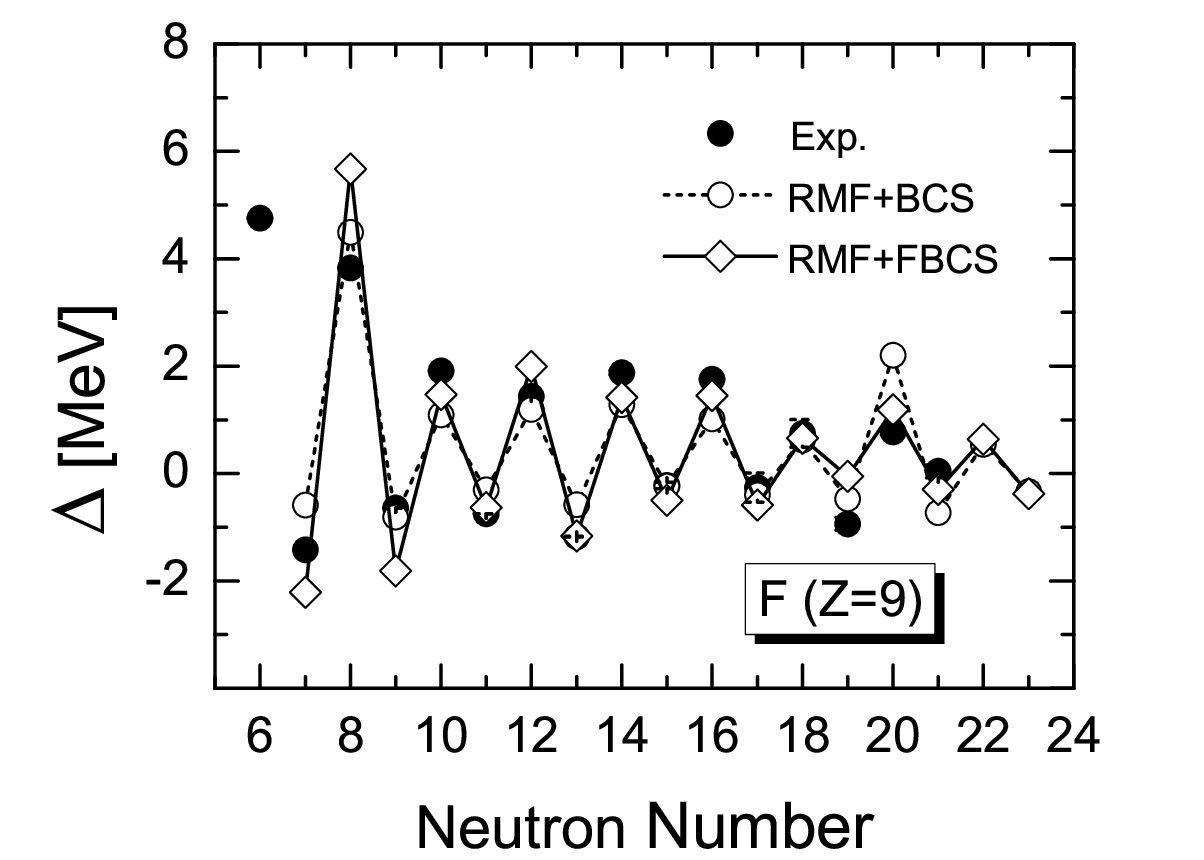}
  \end{minipage}
    \caption{Odd-even mass staggerings of fluorine isotopes obtained by RMF+BCS (open circles) and RMF+FBCS (open diamonds) methods, in comparison with experimental data (solid circles)~\cite{Wangmeng30002}.\label{fig1}}
 \end{center}
\end{figure}

\subsection{Fluorine isotopes}

In Fig.~\ref{fig1}, the odd-even mass staggerings of the fluorine isotopes obtained in
the RMF+BCS and RMF+FBCS methods are compared with the experimental data. Note that
the pairing strengths are fixed by fitting the odd-even mass staggerings of $^{20-24}$F, as shown in Table~2. The difference between the RMF+BCS results and those of the RMF+FBCS becomes larger for
neutron-rich nuclei, which certainly affects the prediction of the drip line.

\begin{figure}[tbp]
 \begin{center}
  \setlength{\abovecaptionskip}{5pt}
  \begin{minipage}[c]{1.0\linewidth}
    \centering
   \includegraphics[scale=0.5]{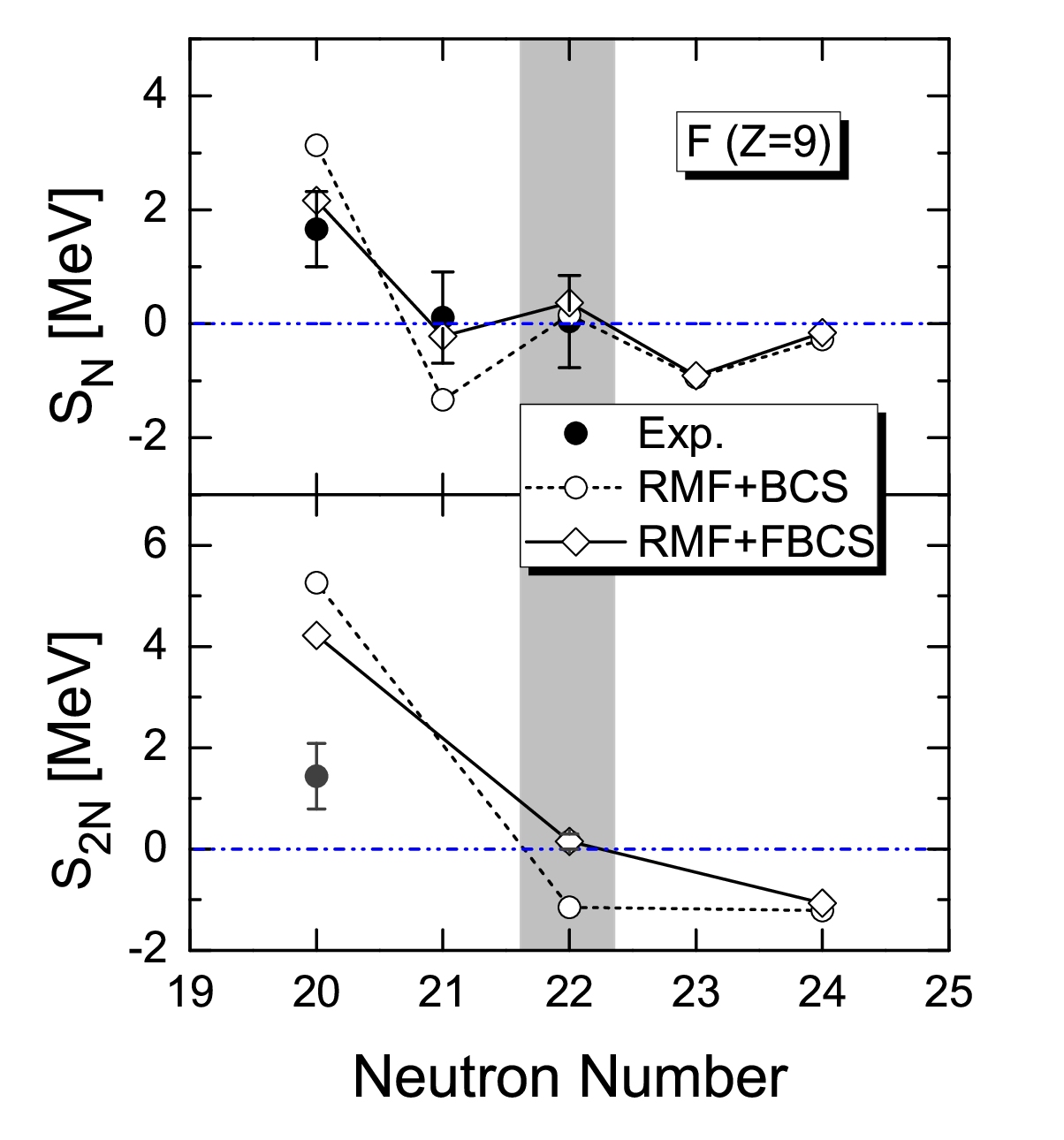}
  \end{minipage}
    \caption{One- (top panel) and two-neutron (bottom panel) separation energies of fluorine isotopes obtained by RMF+BCS (open circles) and RMF+ FBCS (open diamonds) methods, in comparison with experimental data (solid circles)~\cite{Wangmeng30002}. \label{fig2}}
 \end{center}
\end{figure}

In Fig.~\ref{fig2}, the theoretical one- and two-neutron separation energies of the fluorine isotopes
are compared with the experimental data. The two-neutron separation energy is overestimated for $N=20$, similar to  the finite-range droplet model~\cite{PhysRevLett.72.1431}. For the drip-line nucleus, the RMF+BCS approach yields $N=20$.
In contrast, $^{31}$F was experimentally determined as the last bound fluorine
isotope~\cite{PhysRevLett.123.212501}. The two extra nucleons located above the closed $1d_{3/2}$ orbit with $N=22$.
The FBCS method, which restores the particle number conservation, yields $^{31}$F with $N=22$ as the last bound fluorine isotope~\footnote{ To be precise, $^{31}$F is only bound by a few hundreds of keV in the FBCS case. }.

In Ref.~\cite{TANIHATA1995769}, it was argued that if a nucleus is deformed, the Nilsson model is applicable, and
the single-particle states with a given asymptotic quantum number have a degeneracy of two. Then, the drip-line nucleus in the present case would be $N=22$. Therefore, the experiment result~\cite{PhysRevLett.123.212501}
indicates that the neutron-rich fluorine isotopes are deformed. In contrast, although $^{31}$F is strongly deformed in both the RMF+BCS and RFM+FBCS methods (see Fig.~\ref{fig0}), only the latter yields results consistent with the data. This can be attributed to the restored particle number conservation in the FBCS approach. We note that the pairing energy for $^{31}$F in the FBCS method is approximately $2.5~$MeV, whereas it is only $0.2~$MeV in the BCS case.

It was argued in Ref.~\cite{PhysRevC.77.011301} that the restoration of particle
number before variation can stabilize neutron-rich
nuclei, thereby pushing the drip line further away, which agrees with our RMF+FBCS result.

\subsection{Neon isotopes}

\begin{figure}[htbp]
 \begin{center}
  \setlength{\abovecaptionskip}{5pt}
  \begin{minipage}[c]{1.0\linewidth}
    \centering
   \includegraphics[scale=0.5]{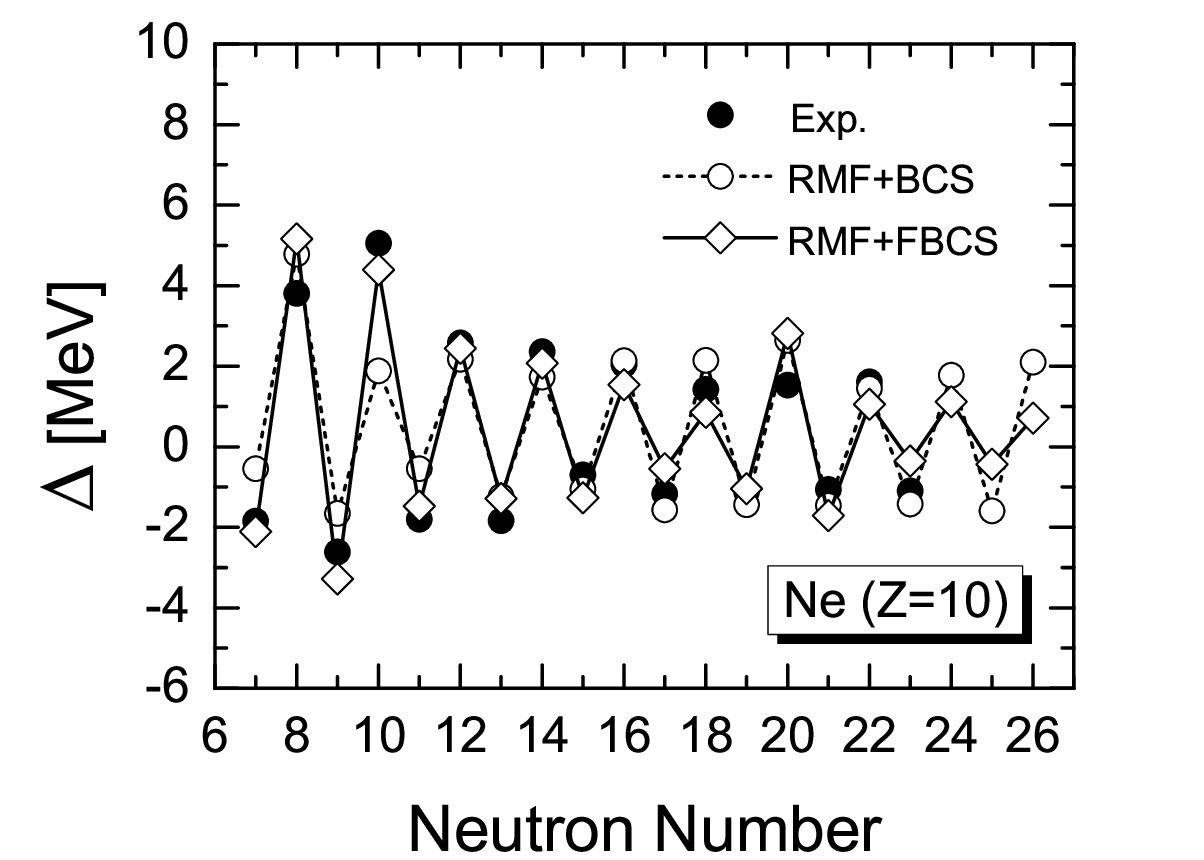}
  \end{minipage}
    \caption{Same as Fig.~\ref{fig1}, for neon isotopes.\label{fig3}}
 \end{center}
\end{figure}

In Fig.~\ref{fig3}, we demonstrate the odd-even mass staggerings of neon isotopes. The theoretical results are
consistent with experimental data, with the exception of
$^{20}$Ne for the BCS method. We note that in the BCS method, only the orbital $2s_{1/2}$ is occupied. However, in the FBCS approach, both $2s_{1/2}$ and $1d_{5/2}$ orbitals are occupied. This explains why the FBCS approach
can better reproduce the empirical pairing gap of $^{20}$Ne.

\begin{figure}[tbp]
 \begin{center}
  \setlength{\abovecaptionskip}{5pt}
  \begin{minipage}[c]{1.0\linewidth}
    \centering
   \includegraphics[scale=0.5]{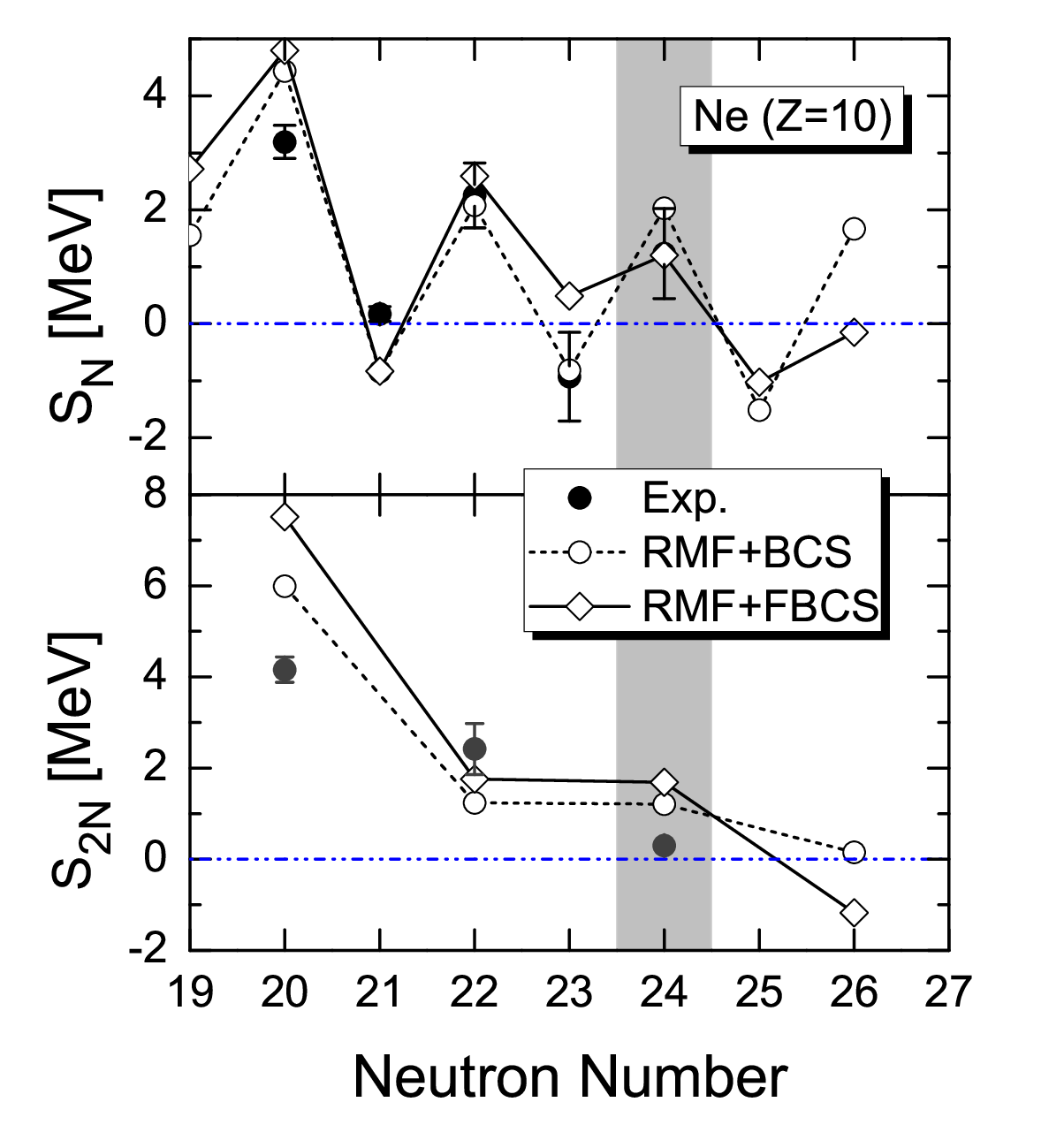}
  \end{minipage}
    \caption{Same as Fig.~\ref{fig2}, for neon isotopes. \label{fig4}}
 \end{center}
\end{figure}

In Fig.~\ref{fig4}, the one- and two-neutron
separation energies of neon isotopes are compared with the experimental data.
The overall agreement is reasonable, with the exception of $^{30}$Ne. For the drip-line nucleus, the BCS method predicts $N=26$,
while the FBCS method predicts $N=24$. The experimental $S_{2N}$ for $^{34}$Ne almost vanishes.
Our FBCS result is qualitatively consistent with the latest experimental data~\cite{PhysRevLett.123.212501}.
The authors of Ref.~\cite{Tanihata:1994wr} also conjectured $^{34}$Ne to be a neutron drip-line nucleus for neon isotopes, the same as the Monte Carlo Shell Model (MCSM) with various particle-hole excitations taken into account~\cite{PhysRevC.60.054315}.

\subsection{Sodium isotopes}

\begin{figure}[htbp]
 \begin{center}
  \setlength{\abovecaptionskip}{5pt}
  \begin{minipage}[c]{1.0\linewidth}
    \centering
   \includegraphics[scale=0.5]{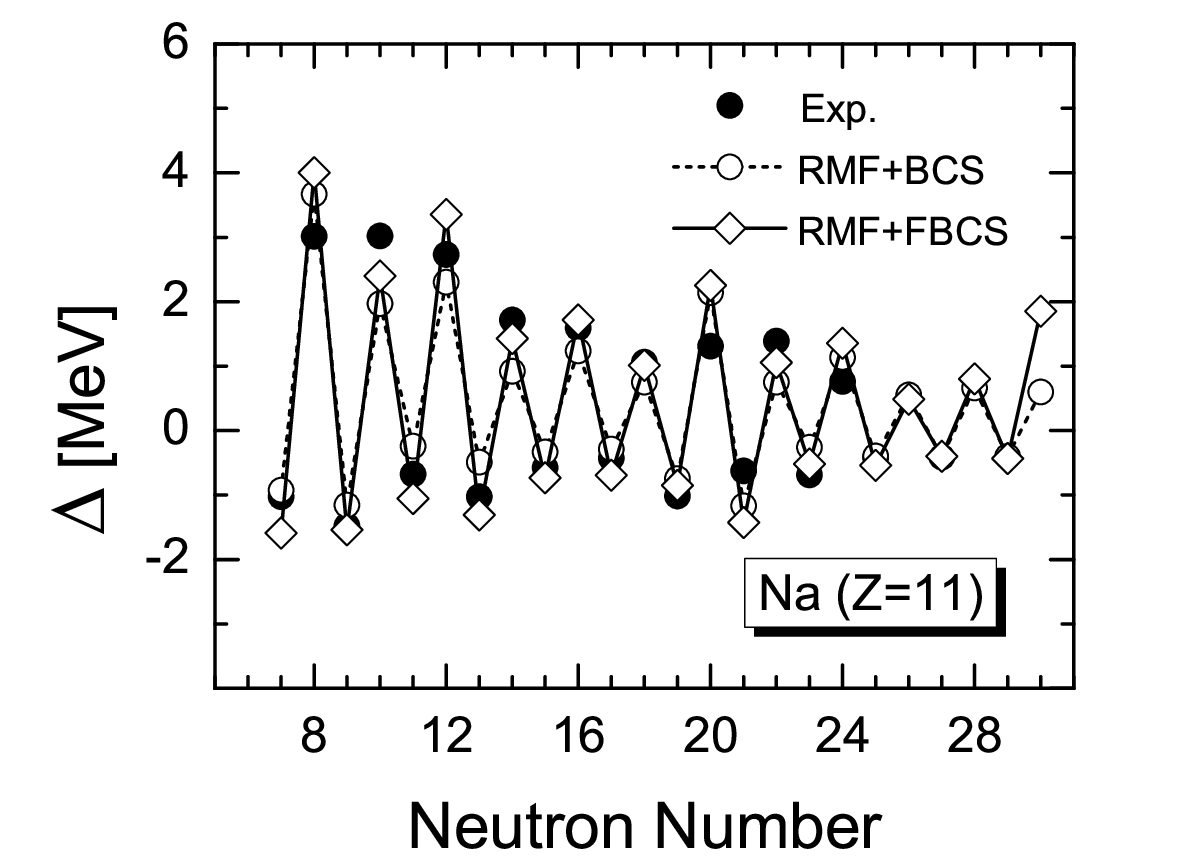}
  \end{minipage}
    \caption{Same as Fig.~\ref{fig1}, for sodium isotopes. \label{fig5}}
 \end{center}
\end{figure}
\begin{figure}[htbp]
 \begin{center}
  \setlength{\abovecaptionskip}{5pt}
  \begin{minipage}[c]{1.0\linewidth}
    \centering
   \includegraphics[scale=0.5]{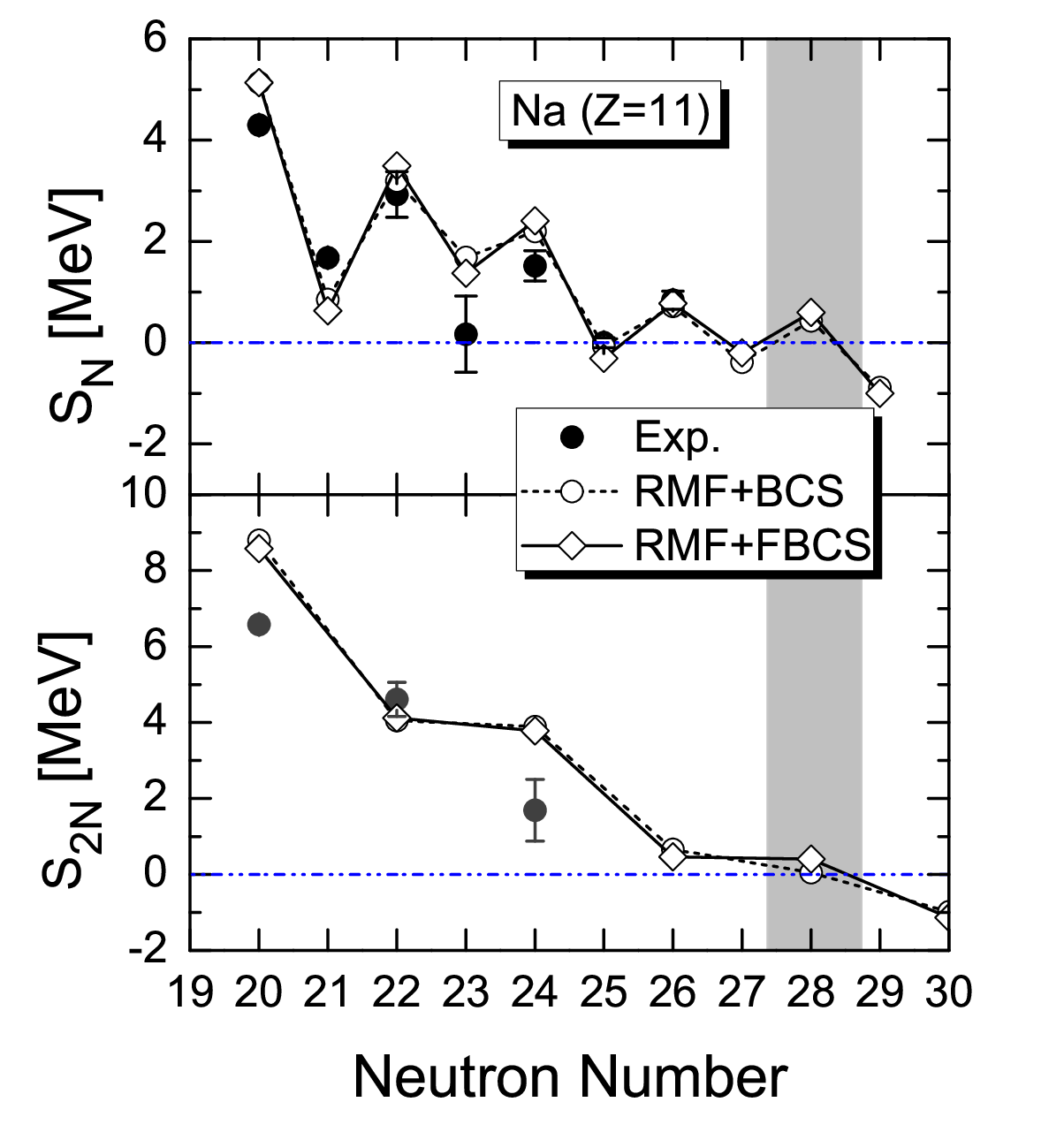}
  \end{minipage}
    \caption{Same as Fig.~\ref{fig2}, for sodium isotopes. \label{fig6}}
 \end{center}
\end{figure}

In Fig.~\ref{fig5}, we show the odd-even mass staggerings, and in Fig.~\ref{fig6},
the one- and two-neutron separation energies of the sodium isotopes in comparison with the experimental data are shown. Interestingly, both the FBCS and BCS methods indicate that $^{39}$Na is the last bound sodium isotope from the perspective of the two-neutron separation energy, though the $S_{2N}$ for $^{39}$Na almost vanishes using the BCS method. Indeed, the recent RIBF experiment detected one event for $^{39}$Na~\cite{PhysRevLett.123.212501}.

The sodium isotopes provide a good opportunity to study the neutron density distributions over a wide range of neutron numbers. In Ref.~\cite{Meng:1998xq}, a systematic study of
nuclear density distributions in the sodium isotopes within the RMF model is discussed, with the
pairing and blocking effect for odd particle systems properly described by the relativistic Hartree-Bogoliubov (HB)
theory in coordinate space. In this study, the neutron drip-line nucleus was predicted
to be $^{45}$Na. As the orbital $1f_{7/2}$ is very close to the continuum, the $N=28$ closed
shell for stable nuclei fails to appear due to the lowering of the $2p_{3/2}$ and $2p_{1/2}$ orbitals.
These results are consistent with the global study using the PC-PK1 parameter set~\cite{Xia:2017zka}. However,
 the influence of deformation is neglected in both studies~\cite{Meng:1998xq,Xia:2017zka}, while the latest RIBF result indicates that deformation effects are important in this part of the nuclear chart.

\subsection{Magnesium isotopes}

\begin{figure}[htbp]
 \begin{center}
  \setlength{\abovecaptionskip}{5pt}
  \begin{minipage}[c]{1.0\linewidth}
    \centering
   \includegraphics[scale=0.5]{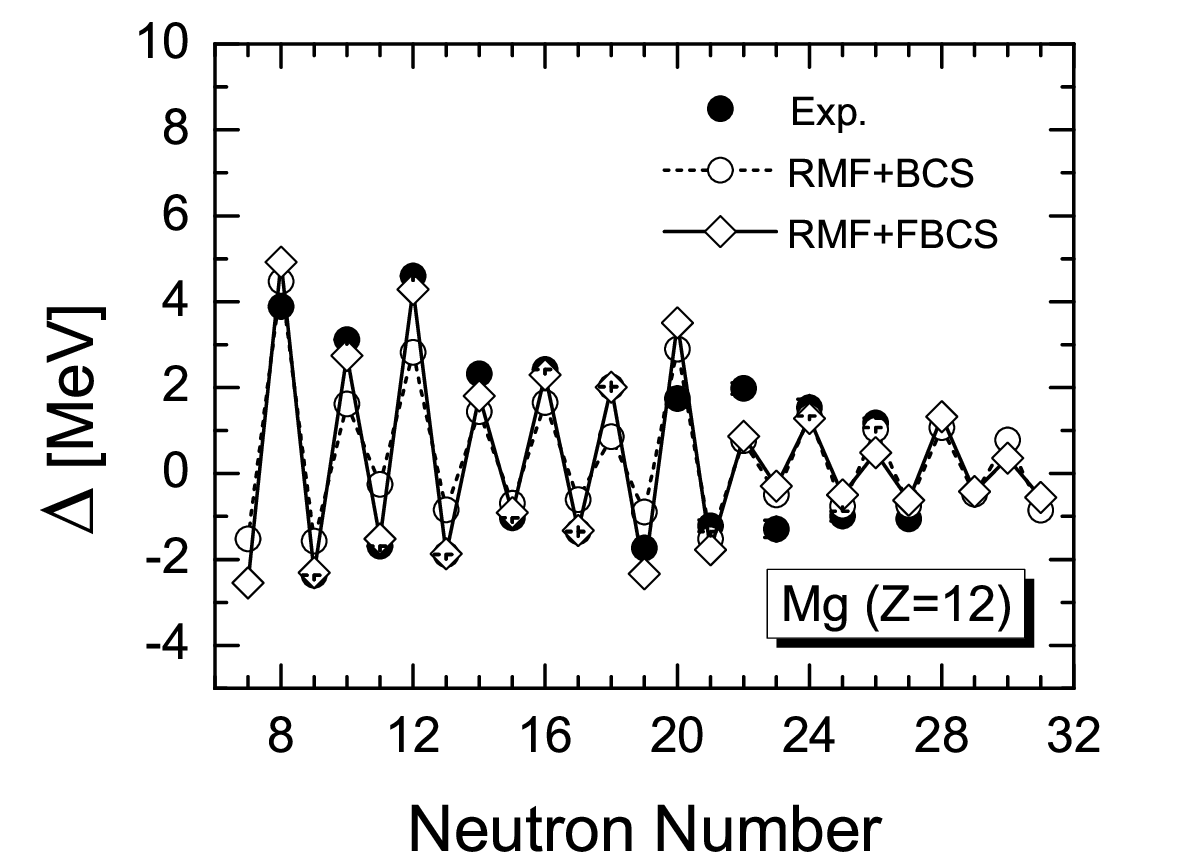}
  \end{minipage}
    \caption{Same as Fig.~\ref{fig1}, for magnesium isotopes. \label{fig7}}
 \end{center}
\end{figure}

\begin{figure}[tbp]
 \begin{center}
  \setlength{\abovecaptionskip}{5pt}
  \begin{minipage}[c]{1.0\linewidth}
    \centering
   \includegraphics[scale=0.5]{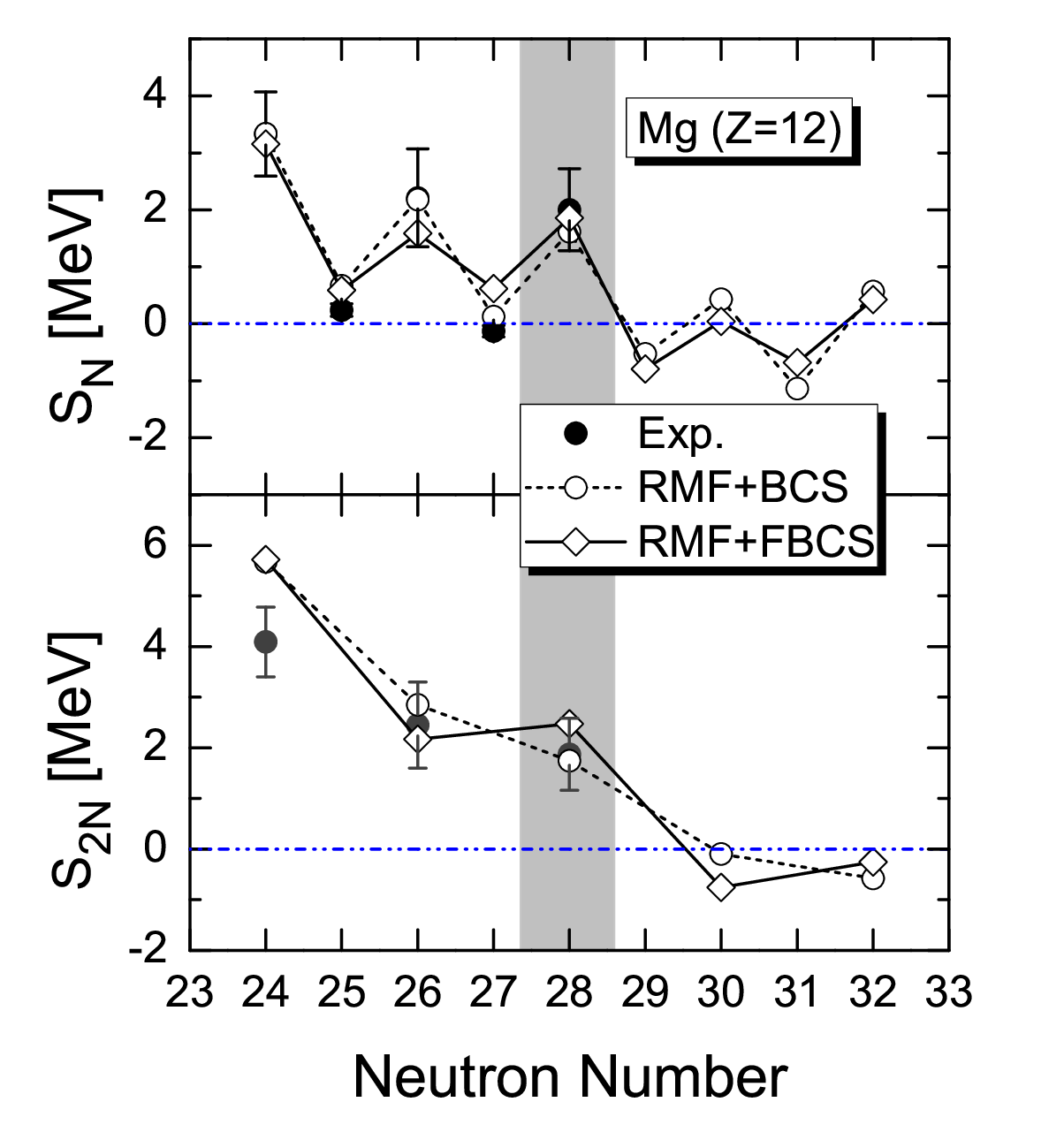}
  \end{minipage}
    \caption{Same as Fig.~\ref{fig2}, for magnesium isotopes. \label{fig8}}
 \end{center}
\end{figure}

The odd-even mass staggerings and one- and two-neutron separation energies of the magnesium isotopes are
shown in Figs.~\ref{fig7} and~\ref{fig8}. From the two-neutron separation energies, we conclude that $^{40}$Mg is the
last bound magnesium isotope. In the BCS approach, $^{42}$Mg is the last bound magnesium
isotope. From Table 1, it is clear that model predictions differ significantly, yielding either $^{40}$Mg, $^{42}$Mg, or $^{46}$Mg as the last bound magnesium isotope. Future experiments are required to settle this issue.

Interestingly, we note from Table 1 that the macroscopic-microscopic mass formula WS4 yields the correct drip-line nuclei $^{24}$O and $^{34}$Ne, except for $^{31}$F. For sodium isotopes, it predicts $^{37}$Na as the last bound neutron-rich nucleus, which is different from our prediction, $^{39}$Na. For the case of magnesium, it predicts $^{40}$Mg as the drip-line nucleus, which is in agreement with the RMF+FBCS approach. Moreover, the WS4 mass formula achieved a description of all existing data at that time with a root-mean-square deviation of 298 keV~\cite{Wang:2014qqa}.


\section{SUMMARY}
Inspired by the latest RIBF experiment that identified the neutron drip line of fluorine and neon isotopes for the first time in twenty years, we studied the impact of particle number conservation in one- and two-neutron separation energies. We employed the recently developed RMF+FBCS approach and studied
fluorine, neon, sodium, and magnesium isotopes. We showed that proper treatment of both the pairing correlations and deformation effects plays an important role in reproducing experimental results, particularly, the
drip-line nuclei of the fluorine and neon isotopes. For the sodium and magnesium isotopes, we predict the
drip-line nuclei to be $^{39}$Na and $^{40}$Mg.

We employed the expansion method based on the harmonic oscillator basis,
which may not be sufficient to account for the continuum effect, though the correct reproduction of the fluorine and neon drip line provide us some confidence in the present study. In the future, the deformed
relativistic Hartree-Bogoliubov theory in continuum (DRHBc) may need to be applied~\cite{Zhou:2009sp,Li:2012gv} to study the impact of the continuum. However, a proper treatment of pairing correlations, particularly from the perspective of particle number conservation, needs to be implemented. In the present work, we adopted the density-independent contact delta interaction in the particle-particle channel; however, we do not anticipate that our conclusion will change if other forms of the
pairing interaction, such as finite range or density-dependent pairing interactions, are considered. Nevertheless, this should be explicitly studied in the future.

As demonstrated in the present study, a proper description of drip-line phenomena is challenging. It requires a proper treatment of numerous effects, including but not limited to the pairing correlation, as well as the deformation and continuum effects. This study shows that the particle number conservation effect can play
a relevant role in identifying the position of the drip line. Further studies are required in the future to corroborate these findings.

\section{Acknowledgements}
This work is partly supported by the National Natural Science Foundation of China under Grant Nos.11735003, 11975041, 11775014, and 11961141004, the fundamental Research Funds for the Central Universities.




\bibliography{refs}
\end{document}